\documentclass[aps,prd,10pt,twocolumn,superscriptaddress,noshowpacs,preprintnumbers,groupedaddress,footinbib,bibnotes]{revtex4-1}

\pdfoutput=1
\usepackage{amsmath}
\usepackage{amsfonts}
\usepackage{amssymb}
\usepackage{mathrsfs}
\usepackage{color}
\usepackage{bm}
\usepackage{graphicx}
\usepackage{blindtext}
\usepackage{wasysym}
\usepackage{mwe}
\usepackage{hyperref}
\usepackage[normalem]{ulem}
\usepackage{blindtext}
\usepackage{syntonly}
\usepackage{cancel}

\newcommand{\ie}{{i.e.}}

\newcommand{\eg}{{e.g.}}

\newcommand{\eq}{Eq.}

\newcommand{\fig}{Fig.}

\newcommand{\Refe}{Ref.}
\newcommand{\Refs}{Refs.}

\newcommand{\equ}[1]{\eq~(\ref{equ:#1})}
\newcommand{\figu}[1]{\fig~\ref{fig:#1}}

\newcommand{\bi}{\begin{itemize}}
\newcommand{\ei}{\end{itemize}}

\begin{document}
 
\title{Bounds on secret neutrino interactions from high-energy astrophysical neutrinos}

\author{Mauricio Bustamante}
\email{mbustamante@nbi.ku.dk}
\thanks{ORCID: \href{http://orcid.org/0000-0001-6923-0865}{0000-0001-6923-0865}}
\affiliation{Niels Bohr International Academy and DARK, Niels Bohr Institute, Blegdamsvej 17, 2100 Copenhagen, Denmark}

\author{Charlotte Rosenstr\o{}m}
\email{vkc652@alumni.ku.dk}
\thanks{ORCID: \href{http://orcid.org/0000-0001-7743-5000}{0000-0001-7743-5000}}
\affiliation{Niels Bohr International Academy and DARK, Niels Bohr Institute, Blegdamsvej 17, 2100 Copenhagen, Denmark}

\author{Shashank Shalgar}
\email{shashank.shalgar@nbi.ku.dk}
\thanks{ORCID: \href{http://orcid.org/0000-0002-2937-6525}{0000-0002-2937-6525}}
\affiliation{Niels Bohr International Academy and DARK, Niels Bohr Institute, Blegdamsvej 17, 2100 Copenhagen, Denmark}

\author{Irene Tamborra}
\email{tamborra@nbi.ku.dk}
\thanks{ORCID: \href{http://orcid.org/0000-0001-7449-104X}{0000-0001-7449-104X}}
\affiliation{Niels Bohr International Academy and DARK, Niels Bohr Institute, Blegdamsvej 17, 2100 Copenhagen, Denmark}

\date{\today}

\begin{abstract}
 Neutrinos offer a window to physics beyond the Standard Model.
 In particular, high-energy astrophysical neutrinos, with TeV--PeV energies, may provide evidence of new, ``secret'' neutrino-neutrino interactions that are stronger than ordinary weak interactions.
 During their propagation over cosmological distances, high-energy neutrinos could interact with the cosmic neutrino background via secret interactions, developing characteristic energy-dependent features in their observed energy distribution.
 For the first time, we use a rigorous statistical analysis to look for signatures of secret neutrino interactions in the diffuse flux of high-energy astrophysical, based on 6 years of publicly available IceCube High Energy Starting Events (HESE).  
 We find no significant evidence for secret neutrino interactions, but place competitive upper limits on the coupling strength of the new mediator through which they occur, in the mediator mass range of 1--100~MeV.
\end{abstract}

\maketitle


\section{Introduction}  

In the Standard Model (SM), because neutrinos interact weakly, neutrino-neutrino interactions are unimportant except in a handful of scenarios with huge neutrino densities, \ie, the early Universe and compact astrophysical objects.  However,  well-motivated proposed extensions of the SM may enhance neutrino-neutrino interactions, rendering them important in other scenarios.  Detecting these {\it secret neutrino interactions} ($\nu$SI) would provide much needed guidance to extend the SM.

\begin{figure}[t!]
 \centering
 \includegraphics[width=\columnwidth]{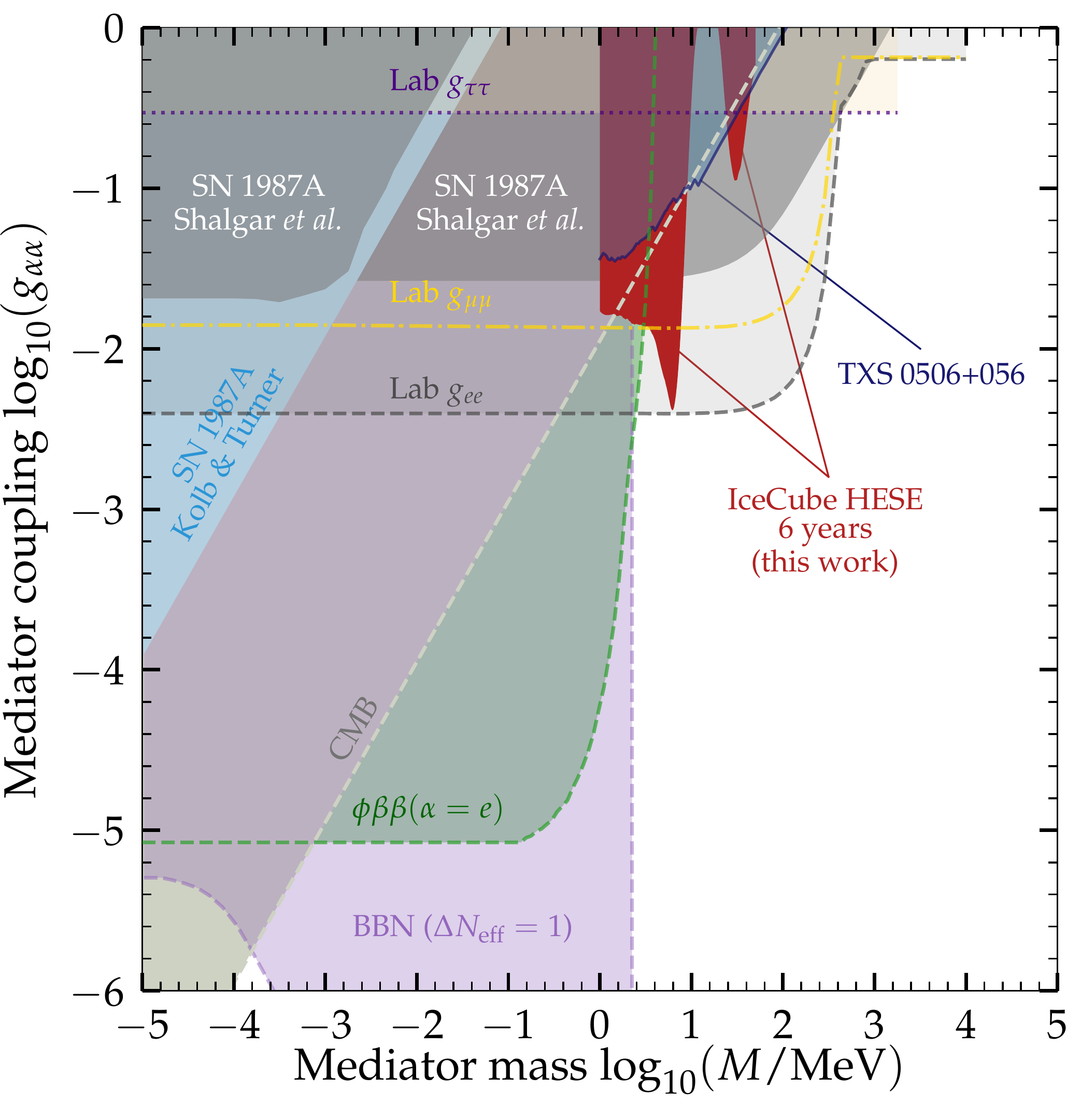}
 \caption{\label{fig:limits}Limits on the mass $M$ and coupling $g_{\alpha\alpha}$ ($\alpha = e, \mu, \tau$) of the new mediator of secret neutrino-neutrino interactions.  Unless otherwise stated, each limit applies to all flavors, \ie, $g_{ee} = g_{\mu\mu} = g_{\tau\tau}$.  Shaded regions are disfavored.  Our limit, at 90\%~C.L., is based on the publicly available 6-year IceCube HESE sample\ \cite{Kopper:2015vzf, IC4yrHESEURL, Kopper:2017zzm}, assuming a neutrino mass of 0.1~eV.  Previous limits come from the CMB\ \cite{Archidiacono:2013dua} (see also \Refe\ \cite{Escudero:2019gvw}), BBN\ \cite{Blinov:2019gcj}, SN 1987A\ \cite{Kolb:1987qy, Shalgar:2019rqe}, particle decays (we distinguish between flavors, \ie, ``Lab $g_{ee}$'', ``Lab $g_{\mu\mu}$'', ``Lab $g_{\tau\tau}$'')\ \cite{Berryman:2018ogk}, and double beta decay ($\phi\beta\beta$, for $\alpha = e$)\ \cite{Brune:2018sab}.  We include the limit estimated from detecting a burst of high-energy neutrinos possibly correlated with the blazar TXS 0506+056, assuming a scalar mediator\ \cite{Kelly:2018tyg}.}
\end{figure}

Secret neutrino interactions occur via a new mediator that predominantly couples to neutrinos; we take their coupling to other particles to be effectively negligible.
The mediator mass $M$ and coupling strength $g$ may be measured using terrestrial experiments or astrophysical observations.
Secret interactions are motivated as solutions to open issues, \eg, the origin of neutrino mass~\cite{Chikashige:1980ui, Gelmini:1980re, Georgi:1981pg, Gelmini:1982rr, Nussinov:1982wu, Blum:2014ewa}, tensions in cosmology~\cite{Aarssen:2012fx, Cherry:2014xra, Barenboim:2019tux, Escudero:2019gvw}, the muon anomalous moment~\cite{Araki:2014ona, Araki:2015mya}, and the LSND anomaly~\cite{Jones:2019tow}.  Presently there is no evidence for $\nu$SI, but \figu{limits} shows that there are strong constraints coming from measurements of the cosmic microwave background (CMB)\ \cite{Cyr-Racine:2013jua, Archidiacono:2013dua, Forastieri:2015paa, Oldengott:2017fhy, Escudero:2019gvw}, Big Bang nucleosynthesis (BBN)\ 
\cite{Ahlgren:2013wba, Huang:2017egl, Blinov:2019gcj}, and the supernova SN 1987A\ \cite{Kolb:1987qy, Shalgar:2019rqe}, where neutrino densities are high; from laboratory measurements of the decay of $Z$ bosons, Higgs bosons, tauons, and $\pi$, $K$, and $D$ mesons\ \cite{Bilenky:1999dn, Lessa:2007up, Laha:2013xua, Berryman:2018ogk}, where branching ratios with final-state neutrinos are precisely measured; and from double beta-decay\ \cite{Agostini:2015nwa, Blum:2018ljv, Brune:2018sab}.  (There are also constraints from supernovae on $\nu$SI via a massless mediator, \eg, \Refs\ \cite{Kachelriess:2000qc, Farzan:2002wx}.)

The high-energy astrophysical neutrinos discovered by the IceCube neutrino telescope, with TeV--PeV energies\ 
\cite{Aartsen:2013bka, Aartsen:2013jdh, Aartsen:2014gkd, Aartsen:2015rwa, Aartsen:2016xlq}, provide a novel, complementary probe of $\nu$SI.  
Owing to their likely extragalactic origin, during their trip to Earth across distances of Mpc--Gpc they may have a significant chance to scatter off the cosmic neutrino background (C$\nu$B) via $\nu$SI.  References\ \cite{Farzan:2014gza, Ng:2014pca, Ioka:2014kca, Ibe:2014pja, Kamada:2015era, DiFranzo:2015qea, Murase:2019xqi} showed that this would introduce characteristic features in the astrophysical neutrino spectrum: a deficit at energies where the scattering is resonant and a pile-up of neutrinos at lower energies.  TeV--PeV neutrinos are especially sensitive to $\nu$SI mediators with masses in the 1--100~MeV range.
(The scattering of neutrinos off extragalactic\ \cite{Farzan:2014gza, Kelly:2018tyg, Pandey:2018wvh, Alvey:2019jzx, Koren:2019wwi, Murase:2019xqi} and Galactic\ \cite{Arguelles:2017atb} dark matter may lead to similar features, but we do not explore that here.)

For the first time, we look for imprints of $\nu$SI in the diffuse flux of high-energy astrophysical neutrinos, using IceCube data.  Our approach is comprehensive: we compute the propagation of astrophysical neutrinos to Earth in the presence of $\nu$SI, followed by their propagation inside Earth, and their detection in IceCube.  We account for uncertainties in the shape of the neutrino spectrum emitted by the sources, atmospheric neutrino backgrounds, and detector uncertainties.

Figure \ref{fig:limits} shows  our results, obtained using the publicly available 6-year sample of IceCube High Energy Starting Events (HESE)\ \cite{Kopper:2015vzf, IC4yrHESEURL, Kopper:2017zzm}.  We find no significant evidence for $\nu$SI, and we place new limits on $M$ and $g$ that overlap with previous ones, providing independent confirmation.  In the absence of knowledge of the precise model that gives rise to $\nu$SI, it is important to probe them using phenomena that involve different energy scales.

This paper is organized as follows.  Section\ \ref{section:astro_nu_intro} gives an overview of high-energy astrophysical neutrinos.  Section\ \ref{section:nuSI} introduces $\nu$SI.  Section\ \ref{section:astro_nu} discusses the propagation of neutrinos under $\nu$SI.  Section\ \ref{section:testing_nusi} shows how we constrain $\nu$SI using IceCube data.  Section\ \ref{section:conclusions} concludes.


\section{High-energy astrophysical neutrinos: an overview}
\label{section:astro_nu_intro}

IceCube has discovered high-energy astrophysical neutrinos\ \cite{Aartsen:2013bka, Aartsen:2013jdh, Aartsen:2014gkd, Aartsen:2015rwa, Aartsen:2016xlq}.  Because they have the highest observed neutrino energies --- up to a few PeV --- and because they likely travel cosmological distances --- up to a few Gpc --- they have a vast potential to explore particle physics at yet-unprobed energy and distance scales\ \cite{Ahlers:2018mkf, Ackermann:2019cxh, Arguelles:2019rbn}.  

The spectrum of high-energy astrophysical neutrinos is fit well by a power law $\propto E_\nu^{-\gamma}$ in the TeV--PeV range.  Using six years of neutrino-induced events that start inside the detector --- the same ones that we use in our analysis --- the IceCube Collaboration found $\gamma = 2.92_{-0.29}^{+0.33}$\ \cite{Kopper:2017zzm}.  (Using instead six years of through-going muon tracks born outside the detector, they found $\gamma \approx 2.13 \pm 0.13$\ \cite{Aartsen:2016xlq}.  However, the two results are compatible within $2\sigma$.)

The origin of the high-energy astrophysical neutrinos is largely unknown, though there are promising candidate astrophysical sources\ \cite{IceCube:2018dnn, IceCube:2018cha, Aartsen:2019fau}.  Because the observed distribution of neutrino arrival directions is isotropic\ \cite{ Ahlers:2015moa, Aartsen:2015knd, Denton:2017csz, Aartsen:2017ujz}, the sources are likely extragalactic.  

Presently, IceCube is the largest neutrino detector.  It consists of an underground array of photomultipliers (PMTs) that instrument a gigaton of ice near the South Pole.  When a high-energy neutrino-nucleon ($\nu N$) interaction (see Section \ref{section:prop_in_earth}) occurs inside or near the instrumented detector volume, it produces high-energy charged particles.  The Cherenkov light that they emit propagates through the ice and the PMTs collect it.  For each detected event, IceCube uses the amount of collected light, and its spatial and temporal profiles, to reconstruct\ \cite{Aartsen:2013vja} the energy $E_{\rm dep}$ deposited by the shower in the detector and the arrival direction $\cos \theta_z$ of the neutrino, where $\theta_z$ is the zenith angle measured from the South Pole.

At TeV--PeV energies, IceCube detects mainly two event topologies: showers and tracks.  Showers are made by the charged-current (CC) $\nu N$ interactions of $\nu_e$ and $\nu_\tau$, and by the neutral-current (NC) $\nu N$ interactions of all flavors.  In a shower, Cherenkov light expands outwards from the interaction vertex with a roughly spherical profile.  Tracks are made by the CC $\nu N$ interactions of $\nu_\mu$, which create, in addition to showers, high-energy muons that travel for a few km and leave tracks of light that are easily identifiable.  Tracks are also made by the CC $\nu N$ interactions of $\nu_\tau$, followed by the decay of the final-state tauon into a muon.  In addition, neutrino-electron interactions contribute to the detection rate, but are largely sub-dominant, except around 6.3~PeV, where $\bar{\nu}_e$ trigger the Glashow resonance\ \cite{Glashow:1960zz}.

Below, we look for evidence of energy-dependent features induced by $\nu$SI on the astrophysical neutrino spectrum using IceCube HESE events\ \cite{Aartsen:2013jdh, Aartsen:2014gkd}, a subset of events where the interaction occurs inside the detector.  
We use HESE events for two reasons.  First, they are of predominantly astrophysical origin.  Below a few tens of TeV, roughly half of them are of astrophysical origin and half are of atmospheric origin\ \cite{Kopper:2017zzm}; above, they are mostly of astrophysical origin\ \cite{Beacom:2004jb, Laha:2013eev}.  This is the result of using the outer layer of PMTs as a self-veto to reduce contamination from atmospheric neutrinos\ 
\cite{Schonert:2008is, Gaisser:2014bja, Arguelles:2018awr}.  Second, in HESE events, most of the neutrino energy is deposited in the detector, which helps preserve the shape of the $\nu$SI-induced spectral features.


\section{Secret neutrino interactions}
\label{section:nuSI}

Secret neutrino interactions are mediated by a new neutral boson that can be a scalar (or pseudo-scalar) or a vector (or axial-vector).  Its mass $M$ and coupling strength $g$ are free parameters; later, we use IceCube data to constrain their values.  

To produce our results, we adopt a scalar mediator $\phi$, so the $\nu$SI interaction term in the flavor basis is $\mathcal{L} \sim g_{\alpha\beta} \phi \bar{\nu}_\alpha \nu_\beta$, where $\alpha, \beta = e, \mu, \tau$.  Neutrinos and anti-neutrinos are equally affected.  We assume that the interaction is flavor-diagonal and universal, \ie, that the only non-zero entries are $g_{\alpha\alpha} \equiv g$.  (See \Refe\ \cite{Kamada:2015era} for an example of a flavor-non-diagonal $\nu$SI in high-energy astrophysical neutrinos.)  In the future, with more statistics, these assumptions could be revisited.  

Because the mediator is a scalar, its decay $\phi \rightarrow \nu + \bar{\nu}$ is helicity-suppressed if neutrinos are Dirac, and is not helicity-suppressed if neutrinos are Majorana.  Hence, the limits that we place below on the $\nu$SI coupling should be understood, in the Dirac case, as applying to the coupling including a helicity-suppression factor, and, in the Majorana case, as applying to the coupling without a helicity-suppression factor.   

Further, our limits, computed for a scalar mediator, can be directly re-interpreted to apply to a vector mediator; in this case, the decay of $\phi$ is not helicity-suppressed for both Dirac and Majorana neutrinos.  Formally, the distribution of neutrino scattering angles in a $\nu$SI interaction is different for scalar and vector mediators.  However, during the propagation to Earth, the final-state relativistic neutrino that emerges from a $\nu$SI interaction is highly boosted in the forward direction.  This reduces any differences in the angular distribution between a scalar and a vector mediator, so that our limits directly apply also to the case of a vector mediator.

We consider the $\nu$SI process $\nu + \bar{\nu} \to \nu + \bar{\nu}$ that may occur during the propagation of astrophysical neutrinos.  One of the initial neutrinos is a high-energy astrophysical neutrino with energy $E_\nu$ in the range TeV--PeV.  The other is a low-energy C$\nu$B neutrino; because its kinetic energy is $\sim$0.1~meV, we take it to be at rest.  In a scattering event, the high-energy neutrino is down-scattered in energy and the low-energy neutrino is up-scattered.  Later, when tracking the propagation of neutrinos, we account for both outcomes (see Section\ \ref{section:astro_nu}).

For the cross section of the above process, we consider only the $s$-channel contribution; it has a resonance, from which our sensitivity to $\nu$SI stems.  The $t$-channel contribution is heavily suppressed\ 
\cite{Ng:2014pca, Farzan:2014gza}.  Thus, we adopt a Breit-Wigner cross section,
\begin{equation}
 \label{equ:sigma}
 \sigma_{\nu\nu}(E)
 =
 \frac{g^4}{16\pi}
 \frac{s}{\left(s-M^2\right)^2+M^2\Gamma^2} \;,
\end{equation}
where $\sqrt{s} \equiv \sqrt{(2 E_\nu m_\nu)}$ is the center-of-mass energy and $\Gamma \equiv g^2 M / (4\pi)$ is the decay width of the mediator; see, \eg, \Refs~\cite{Ioka:2014kca, Ng:2014pca}.  The cross section is resonant at a neutrino energy of $E_{\rm res} = M^2 / (2 m_\nu)$.  
The neutrino mass is unknown; we fix it to $m_\nu = 0.1$~eV, consistent with current constraints\ \cite{Lattanzi:2017ubx}, and ignore small mass differences between flavors.  For $M = 1$--100 MeV, the resonance falls within the TeV--PeV range of the IceCube neutrinos.


\section{Secret interactions in high-energy astrophysical neutrinos}
\label{section:astro_nu}

Below, we compute the diffuse flux of high-energy astrophysical neutrinos, accounting for $\nu$SI on the C$\nu$B and for their propagation inside Earth up to IceCube.  

\vspace*{-0.4cm}


\subsection{Propagating neutrinos to Earth with $\nu$SI}

During the propagation of a high-energy astrophysical neutrino from its source to Earth, it may interact with the C$\nu$B neutrino via $\nu$SI.  The closer the energy of the astrophysical neutrino is to the resonance energy $E_{\rm res}$, and the more distant the source, the higher the chance that $\nu$SI occur.  The cumulative effect of $\nu$SI appears as characteristic spectral features in the energy distribution of high-energy neutrinos at Earth: a dip around $E_{\rm res}$ and a pile-up at lower energies.  

The diffuse flux of high-energy astrophysical neutrinos at Earth is the sum of contributions of neutrinos emitted by all neutrino sources in the local and distant Universe.  Thus, the effect of $\nu$SI is weighted by the redshift evolution of the number density of sources.  We compute the diffuse neutrino flux at Earth following closely the methods from \Refs\ \cite{Farzan:2014gza, Ng:2014pca}.

At time $t$, the comoving number density of one neutrino species of high-energy neutrinos, either $\nu_\alpha$ or $\bar{\nu}_\alpha$, is $n(t)$.  In our convention, neutrinos are emitted at $t < 0$ and reach Earth at $t=0$.  We track the evolution of $\tilde{n}(t, E_\nu) \equiv d n(t,E_\nu) / d E_\nu$ as neutrinos propagate; here, $E_\nu$ is the neutrino energy at time $t$.  Upon reaching Earth, the isotropic diffuse flux of high-energy neutrinos is
\begin{equation*} 
 J_\oplus(E_\nu)
 \equiv
 \frac{d N} {dE_\nu dA dt d\Omega}
 = 
 \frac{c}{4\pi} \tilde{n}(0,E_\nu) \;,
 \vspace*{0.2cm}
\end{equation*}
where $c$ is the speed of light and, to an excellent approximation, the speed of neutrinos.
To compute $\tilde{n}(0,E_\nu)$, we solve the following propagation equation:
\begin{widetext}
 \begin{equation}
  \label{equ:prop_eq}
   \frac {\partial \tilde{n}(t,E_\nu)} {\partial t}
   =
   \frac {\partial} {\partial E_\nu} \left[ b(t,E_\nu) \tilde{n}(t,E_\nu) \right]
   + \mathcal{L}(t, E_\nu)
   - c \, n_{{\rm C}\nu{\rm B}}(t) \sigma_{\nu\nu}(E_\nu) \tilde{n}(t,E_\nu)
   + c \, n_{{\rm C}\nu{\rm B}}(t) \int_{E_\nu}^\infty dE_\nu^\prime \tilde{n}(t,E_\nu^\prime) \frac {d \sigma_{\nu\nu}} {dE_\nu} (E_\nu^\prime) \;.
 \end{equation}
\end{widetext}
To solve \equ{prop_eq}, we first recast it in terms of redshift by using the relation\ \cite{Hogg:1999ad} $dt/dz = -\left[ (1+z) H(z) \right]^{-1}$.  Here, $H(z) \simeq H_0 \sqrt{\Omega_\Lambda + \Omega_m(1+z)^3}$ is the Hubble parameter, $H_0 = 100 h$ km s$^{-1}$ Mpc$^{-1}$ is the Hubble constant, $\Omega_\Lambda$ is the vacuum energy density, and $\Omega_m$ is the matter density.  We fix $h = 0.678$, $\Omega_\Lambda = 0.692$, and $\Omega_m = 0.308$\ \cite{Tanabashi:2018oca}.  Because the interaction is flavor-diagonal and universal, the propagation of all species is computed in the same way, and we need to track the propagation of only one species.  The resulting $\nu$SI-induced spectral features are common to all flavors of $\nu_\alpha$ and $\bar{\nu}_\alpha$.

\begin{figure}[t!]
 \centering
 \includegraphics[width=\columnwidth]{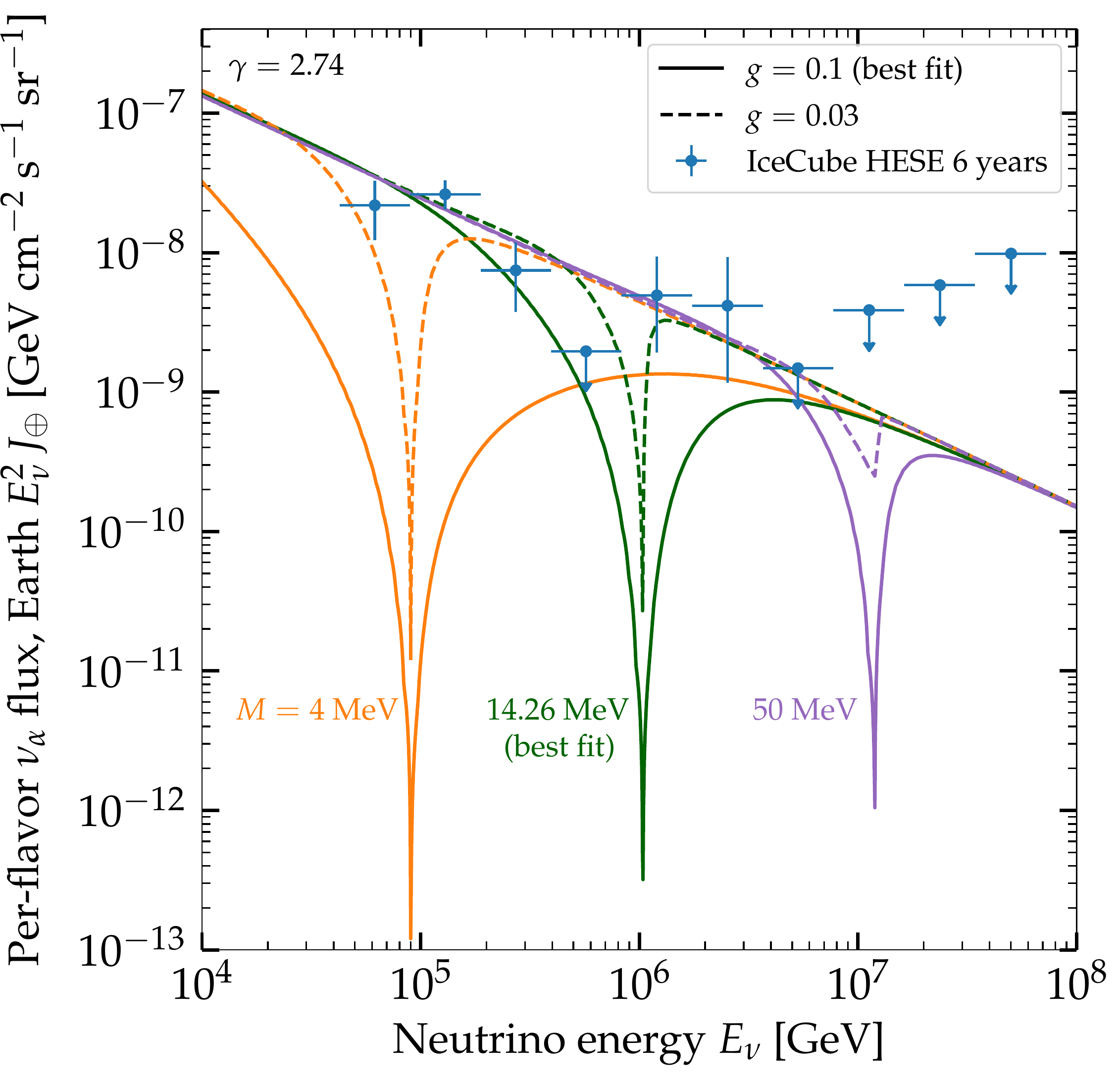}
 \caption{\label{fig:flux_earth}Diffuse per-flavor flux of high-energy astrophysical neutrinos at the surface of the Earth, including the effect of $\nu$SI on the C$\nu$B, for illustrative choices of the mediator mass $M$ and coupling $g$, including their best-fit values from our analysis (see Table\ \ref{tab:fit_results}).  In this plot, the spectral index with which neutrinos are emitted by their sources is fixed to $\gamma = 2.74$ (see Table\ \ref{tab:fit_results}), and the flux is normalized to $E_\nu^2 J_\oplus = 2.46 \cdot 10^{-8}$~GeV~cm$^{-2}$~s$^{-1}$~sr$^{-1}$ at 100~TeV\ \cite{Kopper:2017zzm}, for illustration.}
\end{figure}

In \equ{prop_eq}, the first and second terms on the right-hand side describe the free-streaming of neutrinos, \ie, their propagation in the absence of $\nu$SI.  In this case, the resulting spectrum at Earth would be a pure power law in neutrino energy. 
The first term in \equ{prop_eq} accounts for the continuous energy loss that the neutrinos experience due to the adiabatic cosmological expansion.  The energy loss rate is $b(z,E_\nu) \equiv H(z)E_\nu$.  
The second term in \equ{prop_eq} is the differential number luminosity density of sources, $\mathcal{L}(z, E_\nu) = \mathcal{W}(z) \mathcal{L}_0(E_\nu)$, \ie, the redshift- and energy-dependent injection of neutrinos by the sources; we expand on this below.

The third and fourth terms in \equ{prop_eq} account for $\nu$SI.  
The third term in \equ{prop_eq} accounts for the attenuation of the flux around $E_{\rm res}$.  
The fourth term accounts for the regeneration of neutrinos of initial energy $E_\nu^\prime$ at a new energy $E_\nu$, due to the down-scattering of high-energy astrophysical neutrinos and the up-scattering of low-energy C$\nu$B neutrinos.
The number density of one neutrino species in the C$\nu$B is $n_{{\rm C}\nu{\rm B}}(z) = 56(1+z)^3$~cm$^{-3}$\ \cite{Wong:2011ip}, and the $\nu$SI cross section $\sigma_{\nu\nu}$ is given by \equ{sigma}.

We assume that neutrinos are emitted by a population of extragalactic sources whose number density, $\mathcal{W}$, evolves with redshift following the star formation rate\ \cite{Yuksel:2008cu}, so that most sources lie at $z \approx 1$, corresponding to a distance of a few Gpc.  This assumption holds for promising classes of candidate sources\ \cite{Anchordoqui:2013dnh}.  Following theory expectations, we assume that each source emits neutrinos with a power-law luminosity, \ie, $\mathcal{L}_0(E_\nu) \propto E_\nu^{-\gamma}$. 

We solve \equ{prop_eq} numerically, by integrating from $z_{\max} = 4$ down to $z=0$, with the initial condition $\tilde{n}(z_{\max}, E_\nu) = 0$.  The contribution of sources past $z_{\max}$ is negligible.  Later, when computing fluxes for our analysis (see Section\ \ref{section:testing_nusi}), we treat $M$, $g$, and $\gamma$ as free parameters and let their values be set by a fit to IceCube data.  

Figure\ \ref{fig:flux_earth} shows the diffuse neutrino flux at Earth, for a few illustrative choices of $\gamma$, $M$, and $g$.  Upon reaching Earth, the neutrino spectrum has acquired a deficit, or dip, around $E_{\rm res}$ and a pile-up of down-scattered neutrinos at lower energies.  Although the $\nu$SI cross section, \equ{sigma}, has a sharply defined resonance at $E_{\rm res}$, the dip and pile-up are less sharply defined.  This is because sources emit neutrinos over a relatively wide energy range, not just around $E_{\rm res}$, and because the adiabatic cosmological expansion reduces the energy of neutrinos.

For $M = 1$--100~MeV, the dip and pile-up lie in the TeV--PeV range, where IceCube is sensitive (see Section\ \ref{section:nuSI}).  Therefore, our analysis is sensitive to this mass range.  However, \figu{flux_earth} illustrates that the pile-up is too small to be detected for any value of $M$ and $g$ in the energy range of interest, so our analysis is sensitive only to the existence of the dip.  
Because the decay width of the mediator is $\Gamma \propto g^2 M$, the energy width of the dip grows strongly with the coupling.  For $g \gtrsim 10^{-3}$, the dip is wide and deep, and, in principle, detectable.  For $g \lesssim 10^{-3}$, the dip is shallow and narrow, the spectrum is indistinguishable from the power law expected in the absence of $\nu$SI, and hence our analysis has no sensitivity.

While propagating to Earth, neutrinos change flavor.  We assume that an equal number of astrophysical $\nu_e$, $\nu_\mu$, and $\nu_\tau$ arrives at Earth, in agreement with expectations from standard flavor-mixing\ 
\cite{Beacom:2003nh, Kashti:2005qa, Lipari:2007su, Mena:2014sja, Bustamante:2015waa, Bustamante:2019sdb} and with IceCube results\ \cite{Aartsen:2015knd, Aartsen:2015ivb}.  We also assume equal fluxes of neutrinos and anti-neutrinos, since presently they are indistinguishable in IceCube.  Equal neutrino and anti-neutrino fluxes are expected, for instance, from neutrino production via proton-proton interactions\ \cite{Stecker:1978ah, Kelner:2006tc}.


\subsection{Propagating neutrinos inside the Earth}  
\label{section:prop_in_earth}

\begin{figure}[t!]
 \centering
 \includegraphics[width=\columnwidth]{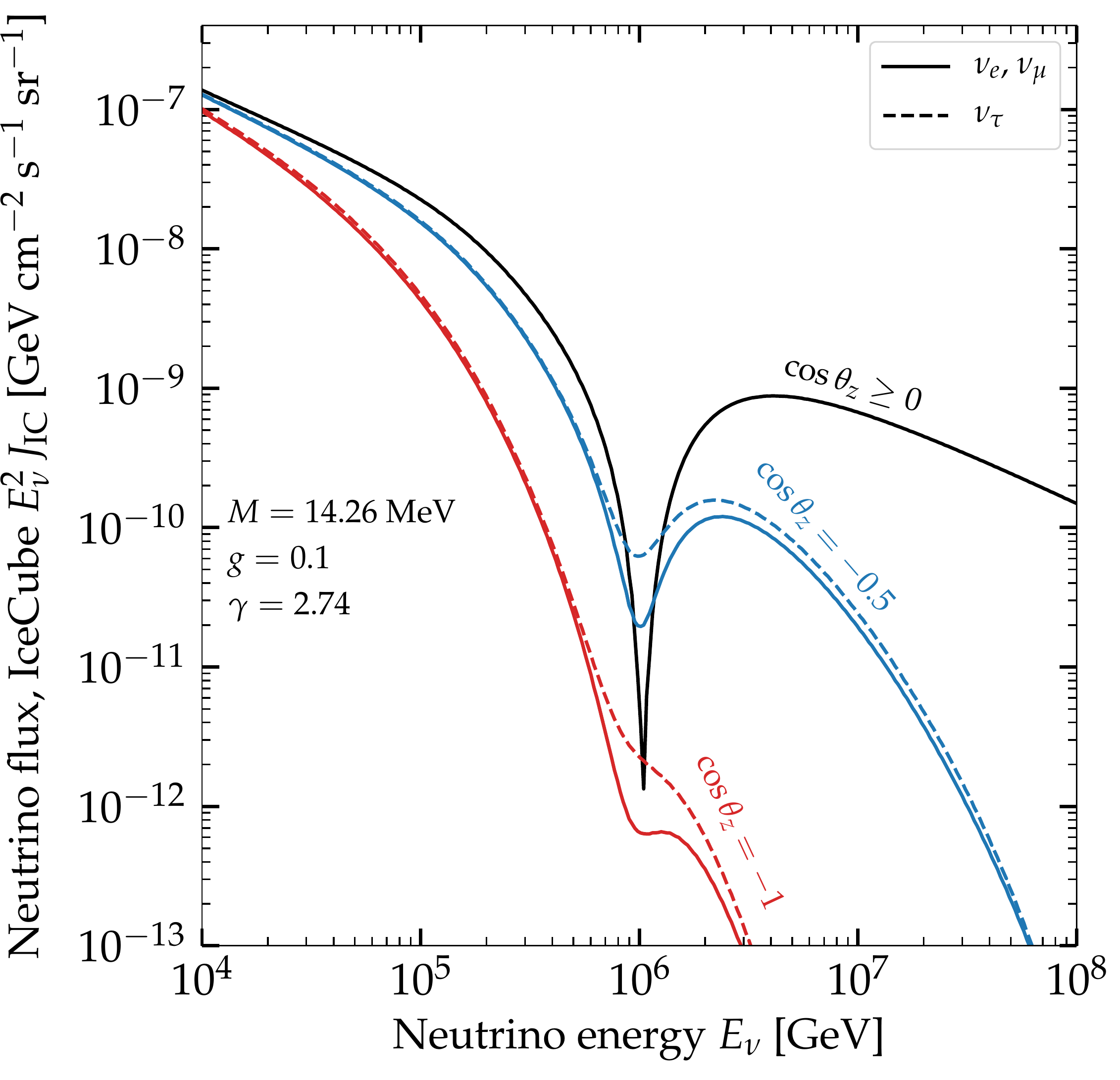}
 \caption{\label{fig:flux_detector}Diffuse per-flavor flux of high-energy astrophysical neutrinos that reach IceCube after propagating inside Earth along illustrative directions: from above ($\cos \theta_z \geq 0$), from $60^\circ$ below the horizon ($\cos \theta_z = -0.5$), and directly from below ($\cos \theta_z = -1$).  In this plot, the fluxes have the same normalization as in \figu{flux_earth}, and $M$, $g$, and $\gamma$ are set to their best-fit values (see Table\ \ref{tab:fit_results}).  For $\cos \theta_z \geq 0$, the effect of in-Earth propagation is negligible, and  the curves for all flavors overlap.  Results for $\bar{\nu}_e$, $\bar{\nu}_\mu$, and $\bar{\nu}_\tau$ (not shown) are similar.}
 \vspace*{-0.3cm}
\end{figure}

Neutrinos with energies above 10~TeV have a significant chance of scattering off nucleons as they propagate inside the Earth.  The longer their path inside the Earth, the higher the chance that they scatter\ \cite{Aartsen:2017kpd, Bustamante:2017xuy}.  As a result, while the flux of astrophysical neutrinos is isotropic at the surface of the Earth, the flux that arrives at IceCube, after traveling inside the Earth along different directions, is no longer isotropic.

At these energies, a neutrino typically interacts with a nucleon $N$ via deep inelastic scattering\ \cite{Conrad:1997ne, Formaggio:2013kya}: the neutrino scatters off the partons of the nucleon and breaks it up.  The CC channel of this interaction attenuates the flux by removing neutrinos, \ie, $\nu_\alpha + N \to \alpha + X$, where $X$ are final-state hadrons.  In the case of $\nu_\tau$, a CC interaction produces a tau that propagates for some distance before decaying again into a $\nu_\tau$; as a result of this regeneration, the flux of $\nu_\tau$ that reaches IceCube is less attenuated than that of $\nu_e$ and $\nu_\mu$.  The NC channel dampens the energy of neutrinos, \ie, $\nu_\alpha + N \to \nu_\alpha + X$, where the final-state neutrino carries, on average, 70\% of the energy of the parent neutrino\ \cite{Gandhi:1995tf}.  

We propagate all flavors of astrophysical and atmospheric (see below) neutrinos through the Earth and up to IceCube, in the direction of the HESE events, using {\tt nuSQuIDS}\ \cite{Delgado:2014kpa, SQuIDS, NuSQuIDS}, which takes into account the aforementioned interactions.  We propagate $\nu_\alpha$ and $\bar{\nu}_\alpha$ separately, since the cross section for the latter is up to 50\% smaller at a few tens of TeV\ \cite{Connolly:2011vc, CooperSarkar:2011pa}.  For the matter density inside the Earth, we adopt the Preliminary Reference Earth Model\ \cite{Dziewonski:1981xy}, and assume that the matter is iso-scalar, \ie, that it is made up of protons and neutrons in equal proportions.  At these high energies, there are no matter-driven flavor transitions.

Figure\ \ref{fig:flux_detector} illustrates the effect of in-Earth propagation on the astrophysical neutrino flux, for a particular choice of values of $\nu$SI parameters and for different arrival directions at IceCube.  The propagation shifts the $\nu$SI dip to slightly lower energies and widens it.  For downgoing neutrinos ($\cos \theta_z \geq 0$), the path length inside Earth is small and the effect of $\nu N$ interactions is negligible.  For upgoing neutrinos ($\cos \theta_z < 0$), the effect is significant and grows with the path length.  The $\nu_\tau$ flux is less affected due to its regeneration inside the Earth.  Below, as part of our analysis, we compute expected neutrino-induced event rates at IceCube along different arrival directions; when doing so, we always propagate first the flux inside the Earth up to IceCube (see Section\ \ref{section:testing_nusi}).


\section{Testing for $\nu$SI}
\label{section:testing_nusi}

We look for the presence of the $\nu$SI-induced spectral features described in Section\ \ref{section:astro_nu} in the publicly available 6-year IceCube HESE sample, consisting of 80 events, 58 showers and 22 tracks, with deposited energies between 18~TeV and 2~PeV\ 
\cite{Kopper:2015vzf, IC4yrHESEURL, Kopper:2017zzm}.  For each event, $E_{\rm dep}$ and $\cos \theta_z$ are provided.  
In the sample, there are no events in the range $E_{\rm dep} \approx 300$~TeV--1~PeV.  Later, we show that this gap in events impacts our results significantly.


\subsection{Computing HESE event rates}

For a given neutrino flux that arrives at the detector, we compute the HESE event rate following the detailed procedure introduced in \Refe\ \cite{Palomares-Ruiz:2015mka}.  In it, the energy $E_\nu$ of the interacting neutrino is converted into electromagnetically equivalent deposited energy $E_{\rm dep}$.  This is the energy that is ultimately registered as Cherenkov light.  The conversion differs for tracks, hadronic showers --- initiated by final-state hadrons --- and electromagnetic showers --- initiated by electrons.  We compute detection via the dominant neutrino-nucleon interactions and the sub-dominant neutrino-electron interactions.  Further, we include a 12\% detector energy resolution on $E_{\rm dep}$.  Appendix\ \ref{appendix:detection} sketches the procedure; for details, see \Refe\ \cite{Palomares-Ruiz:2015mka}.


\subsection{Atmospheric neutrino and muon backgrounds}

High-energy atmospheric neutrinos and muons born in cosmic-ray interactions in the atmosphere are the dominant background in searches for high-energy astrophysical neutrinos.  In our analysis, we account in detail for their contribution to the IceCube event rate.  

For atmospheric neutrinos, we use the same state-of-the-art tools used by the IceCube Collaboration: {\tt MCEq} to compute fluxes at the surface of the Earth\ \cite{Fedynitch:2015zma, MCEq} and {\tt nuVeto} to compute the HESE self-veto\ \cite{Arguelles:2018awr, nuVeto}.  We consider only the contribution of conventional atmospheric neutrinos, from the decay of pions and kaons, and neglect the contribution of prompt atmospheric neutrinos, from the decay of charmed mesons, since they remain undiscovered and subject to severe upper limits\ \cite{Aartsen:2015knd}.   

For atmospheric muons, we approximate the flux that reaches IceCube following the approach of \Refe\ \cite{Palomares-Ruiz:2015mka}, which is based on measurements.  Appendix\ \ref{appendix:atm_bg} contains details about the atmospheric backgrounds.


\subsection{Statistical analysis}

To look for evidence of $\nu$SI in the IceCube data, we generate test HESE samples, following Appendix\ \ref{appendix:detection}, with varying values of $M$, $g$, and $\gamma$.   Then we compare the test samples to the observed 6-year IceCube HESE sample\ \cite{Kopper:2015vzf, IC4yrHESEURL, Kopper:2017zzm}.  We adopt a Bayesian approach and perform the comparison by maximizing a likelihood function.  Our statistical method is modeled after \Refs\ \cite{Palomares-Ruiz:2015mka, Vincent:2016nut, Bustamante:2017xuy}.

\begin{table}[t!]
 \begin{ruledtabular}
  \caption{\label{tab:fit_results}Mass $M$ and coupling strength $g$ of the new $\nu$SI mediator obtained in our statistical analysis, using the publicly available 6-year sample of IceCube HESE events.  The allowed range of values for each parameter is marginalized over all of the remaining parameters.
  The other parameters obtained in the same statistical analysis, \ie, with $\nu$SI, are $\gamma$, the spectral index with which astrophysical neutrinos are emitted; and $N_{\rm ast}$, $N_{\rm atm}$, and $N_\mu$, the number of astrophysical neutrinos, conventional atmospheric neutrinos, and atmospheric muons in the sample, respectively.  See the main text for details.}
  \centering
  \begin{tabular}{cccc}
   Parameter & Best fit $\pm 1\sigma$ & $2\sigma$ & $3\sigma$ \\
   \hline
   $\log_{10}(M/{\rm MeV})$  & $1.154 \pm 0.073$        & $[0.56, 2.53]$    & $\leq 2.97$   \\
   $M$~[MeV]                 & $14.26_{-2.21}^{+2.61}$  & $[3.63, 338.84]$  & $\leq 933.25$ \\
   $\log_{10} g$             & $-1.00 \pm 0.97$         & $[-6.03, -0.73]$  & $\leq -0.30$  \\
   $g$                       & $0.1_{-0.09}^{+0.83}$    & $[9 \cdot 10^{-6}, 0.19]$  & $\leq 0.50$  \\
   \hline
   $\gamma$        & $2.74 \pm 0.13$ &   \\
   $N_{\rm ast}$   & $64.7 \pm 6.2$ &   \\
   $N_{\rm atm}$   & $15.1 \pm 2.6$ &   \\
   $N_\mu$         & $7.6 \pm 3.8$ &  \\
  \end{tabular}
 \end{ruledtabular}
\end{table}

For the 6-year HESE data set, which contains $N_{\rm obs} = 80$ events, the likelihood function is
\begin{eqnarray}
 \label{equ:full_likelihood}
 && \mathcal{L}(M, g, \gamma, N_{\rm ast}, N_{\rm atm}, N_\mu) 
 =
 e^{-N_{\rm ast}-N_{\rm atm}-N_\mu}
 \nonumber \\
 && 
 \qquad \qquad \quad
 \times
 \prod_{i=1}^{N_{\rm obs}} \mathcal{L}_i(M, g, \gamma, N_{\rm ast}, N_{\rm atm}, N_\mu) \;,
\end{eqnarray}
where $N_{\rm ast}$, $N_{\rm atm}$, and $N_\mu$ are, respectively, the number of HESE events due to astrophysical neutrinos, conventional atmospheric neutrinos, and atmospheric muons. 

The partial likelihood for the $i$-th event compares the chances of it being due to the different fluxes, \ie,
\begin{eqnarray}
 && \mathcal{L}_i(M, g, \gamma, N_{\rm ast}, N_{\rm atm}, N_\mu)
 \nonumber \\
 &&
 \qquad
 =
 N_{\rm ast} \mathcal{P}_{i, {\rm ast}}(M, g, \gamma)
 +
 N_{\rm atm} \mathcal{P}_{i, {\rm atm}} 
 +
 N_{\mu} \mathcal{P}_{i, \mu} \;,
\end{eqnarray}
where $\mathcal{P}_{i, {\rm ast}}$, $\mathcal{P}_{i, {\rm atm}}$, and $\mathcal{P}_{i, \mu}$ are, respectively, the probability distribution functions for this event to have been generated by the flux of astrophysical neutrinos, atmospheric neutrinos, and atmospheric muons.  For astrophysical neutrinos, this is 
\begin{equation}
 \label{equ:pdf_ast}
 \mathcal{P}_{i, {\rm ast}}(M, g, \gamma)
 =
 \frac
 {
 \left.
 \frac{dN^{\rm ast}(M, g, \gamma)}{dE_{\rm dep}}
 \right\vert_{E_{{\rm dep}, i}, \cos \theta_{z, i}} 
 }
 {
 \int_{E_{\rm dep}^{\min}}^{E_{\rm dep}^{\max}} dE_{\rm dep}
 \left.
 \frac{dN^{\rm ast}(M, g, \gamma)}{dE_{\rm dep}}
 \right\vert_{\cos \theta_{z, i}}
 }
 \;,
\end{equation} 
where $E_{\rm dep}^{\min} = 10^4$~GeV, $E_{\rm dep}^{\max} = 10^7$~GeV, and $E_{{\rm dep}, i}$ and $\cos \theta_{z,i}$ are the deposited energy and direction of the event.  The event spectrum $dN^{\rm ast} / dE_{\rm dep}$ is given by \equ{spectrum_total_sh}, if the event is a shower, or by \equ{spectrum_total_tr}, if it is a track.  Analogously, for atmospheric neutrinos, 
\begin{equation}
 \label{equ:pdf_conv}
 \mathcal{P}_{i, {\rm atm}}
 =
 \frac
 {
 \left.
 \frac{dN^{\rm atm}}{dE_{\rm dep}}
 \right\vert_{E_{{\rm dep}, i}, \cos \theta_{z, i}} 
 }
 {
 \int_{E_{\rm dep}^{\min}}^{E_{\rm dep}^{\max}} dE_{\rm dep}
 \left.
 \frac{dN^{\rm atm}}{dE_{\rm dep}}
 \right\vert_{\cos \theta_{z, i}}
 }
 \;,
\end{equation} 
and the event spectrum $dN^{\rm atm}/dE_{\rm dep}$ is given by \equ{spectrum_total_sh} or \equ{spectrum_total_tr} if the event is a shower or a track, respectively.  Because atmospheric muons only contribute to the rate of tracks, the probability distribution function is the same as \equ{pdf_conv}, but with $dN^{\rm atm}/dE_{\rm dep} \to  dN^{{\rm tr}, \mu}/dE_{\rm dep}$, as described in Appendix\ \ref{appendix:detection}.

The likelihood, \equ{full_likelihood}, depends on 6 free parameters: $M$, $g$, $\gamma$, $N_{\rm ast}$, $N_{\rm atm}$, and $N_\mu$.  
When maximizing it, we avoid introducing unnecessary bias by choosing generous flat priors for the physical parameters, in log space for the $\nu$SI parameters and in linear space for the astrophysical spectral index: 
$\log_{10} (M/{\rm MeV}) \in [-1, 3]$, $\log_{10} g \in [-7,0]$, and $\gamma \in [2,3]$. 
For $N_{\rm ast}$, we also choose a flat prior: $N_{\rm ast} \in [0,N_{\rm obs}]$.  
For $N_{\rm atm}$ and $N_\mu$, we choose priors following the expected contribution of atmospheric neutrinos and muons to the 6-year IceCube HESE sample\ \cite{Kopper:2017zzm}.  
For $N_{\rm atm}$, we choose a skew normal prior with central value and asymmetric errors of $15.6_{-3.9}^{+11.4}$.
For $N_\mu$, we choose a normal prior with central value and symmetric errors of $25.2 \pm 7.3$.

To maximize the likelihood, we use {\tt MultiNest}\ \cite{Feroz:2007kg, Feroz:2008xx, Feroz:2013hea, Buchner:2014nha}, an efficient implementation of the multimodal importance nested sampling algorithm for Bayesian analysis.  When reporting two-dimensional marginalized contours in \figu{best_fit}, we use {\tt GetDist}\ \cite{Lewis:2019xzd, GetDist}.


\subsection{Results}  

\begin{figure}[t!]
 \centering
 \includegraphics[width=0.965\columnwidth]{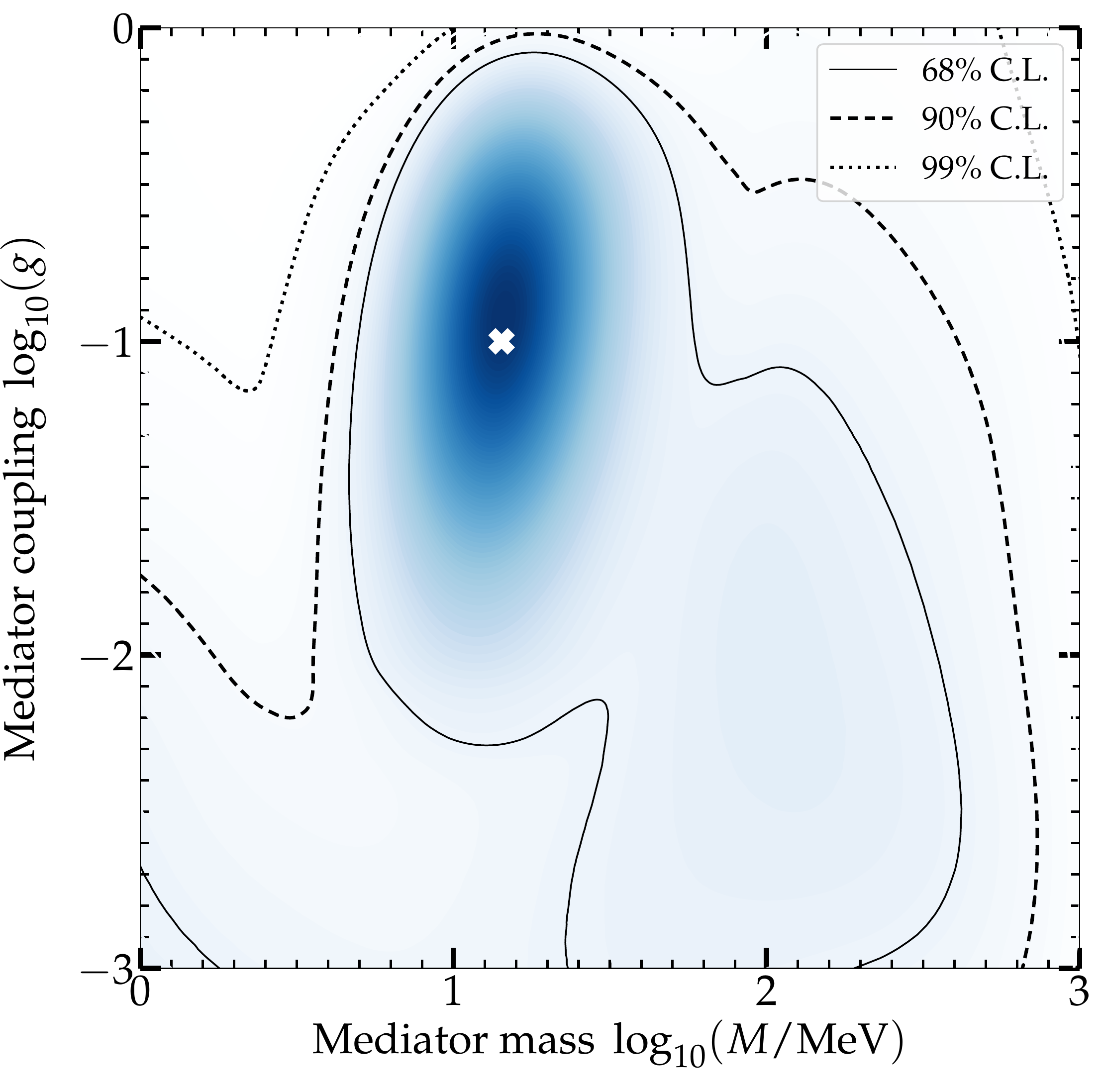}
 \caption{\label{fig:best_fit}Best-fit values (white marker) and two-dimensional marginalized credible regions of the mass $M$ and coupling $g$ of the $\nu$SI mediator, resulting from our fit to the 6-year IceCube HESE data sample.  The shading represents the posterior probability density.  See Table\ \ref{tab:fit_results} for the one-dimensional marginalized values and the main text for details.}
\end{figure}

Table\ \ref{tab:fit_results} shows the resulting one-dimensional marginalized allowed ranges of all likelihood parameters.  Our fit value of $\gamma = 2.74 \pm 0.13$, allowing for $\nu$SI, is compatible within $1\sigma$ with the value reported by the IceCube Collaboration without $\nu$SI, \ie, $\gamma = 2.92_{-0.29}^{+0.33}$\ \cite{Kopper:2017zzm}.  For the $\nu$SI parameters, the fit yields $M = 14.26_{-2.21}^{+2.61}$~MeV and $g = 0.1_{-0.09}^{+0.83}$.  At $1\sigma$, the range of $M$ is narrow, but the range of $g$ is wide.  At $2\sigma$ already, tiny values of $g$ are allowed, for which no $\nu$SI-induced spectral features would be discernible.  At $3\sigma$, $g$ is essentially unconstrained.

Thus, we find no statistically significant evidence for $\nu$SI in the 6-year HESE sample.  
The Bayes factor comparing the Bayesian evidence of our fit to a fit that does not allow for $\nu$SI is $\ln B \approx 2.48$.  In Jeffreys' scale\ \cite{Jeffreys:1939xee}, this falls short of ``moderate evidence''.  The corresponding frequentist significance\ \cite{Trotta:2008qt}, or p-value, is $\sim$0.006, equivalent to $\sim$$2.7\sigma$, insufficient to claim discovery of $\nu$SI.

Figure\ \ref{fig:best_fit} shows that the marginalized allowed regions in the $g$~{\it vs.}~$M$ parameter space are large, supporting our earlier remarks.
Centered around $\log_{10} (M/{\rm MeV}) = 2$, the contours exhibit a shoulder-like feature, albeit with a very low posterior probability density.  This feature is driven by the lack of HESE events beyond 2~PeV, which the fit attempts to attribute to the existence of a $\nu$SI-induced spectral dip (see also \Refe\ \cite{Mohanty:2018cmq}).  Its posterior is low because there is no narrow event gap for the dip to fit, but rather just an absence of multi-PeV events.  

\begin{figure}[t!]
 \centering
 \includegraphics[width=\columnwidth]{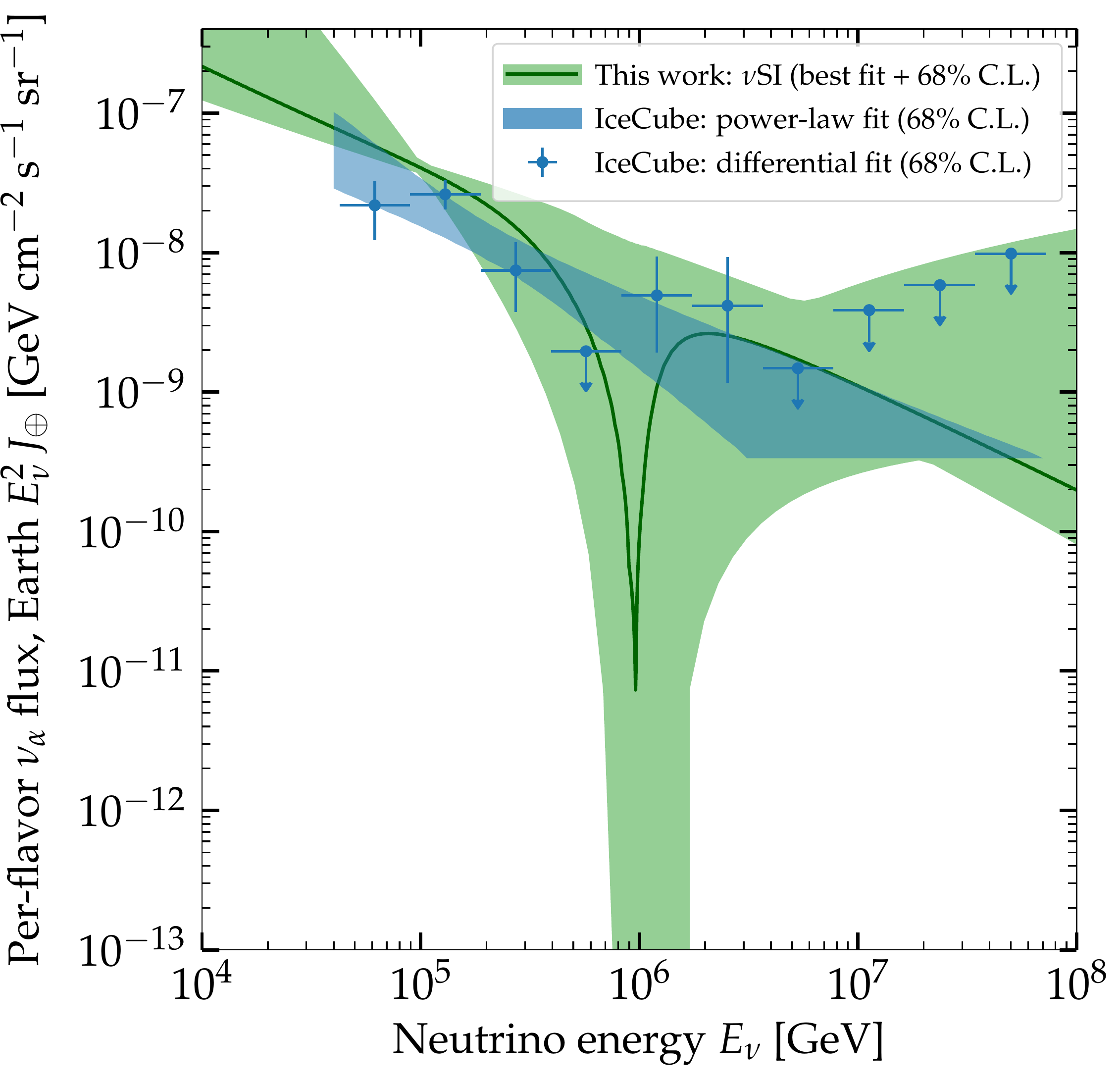}
 \vspace*{-0.5cm}
 \caption{\label{fig:flux_best_fit}Diffuse flux of high-energy astrophysical neutrinos at the surface of the Earth, including the effect of $\nu$SI interactions, computed by varying the parameters $M$, $g$, $\gamma$, and $N_{\rm ast}$ within their $1\sigma$ ranges from Table\ \ref{tab:fit_results}.  For comparison, we include the differential and power-law fits to the 6-year HESE sample as reported by IceCube\ \cite{Kopper:2017zzm}.}
\end{figure}

Figure\ \ref{fig:flux_best_fit} shows that the best-fit values of $M$, $g$, and $\gamma$ yield an astrophysical neutrino flux with a wide $\nu$SI-induced spectral dip centered around 1~PeV.  The dip roughly fills the event gap that exists in the 6-year HESE sample in the range $E_{\rm dep} \approx 300$~TeV--1~PeV.   Further, \figu{flux_best_fit} shows that the region of neutrino fluxes with $\nu$SI generated by varying $M$, $g$, $\gamma$, and $N_{\rm ast}$ within their allowed ranges from Table\ \ref{tab:fit_results} is compatible with the region generated by fitting a pure power-law spectrum, as reported by the IceCube Collaboration\ \cite{Kopper:2017zzm}.

Figure\ \ref{fig:limits} shows our upper limits on $g$ as a function of $M$.  Our limits cover roughly the range $M = 1$--100~MeV.  There, they overlap with existing limits and provide independent confirmation.  Our limits are strongest around $M \approx 6$~MeV, which corresponds to a resonant neutrino energy of $E_{\rm res} \approx 200$~TeV, close to the energy range where the gap in the HESE events is.  Our limits are weakest around the best-fit value of $M = 14.26$~MeV, as expected.  
Similarly to the shoulder region in \figu{best_fit}, the small disconnected region of weak limits around $M = 30$~MeV in \figu{limits} corresponds to $\nu$SI-induced spectral dips centered at energies of 2~PeV and above, where there are no more events in the sample.


\subsection{Comparison with previous work}  

References \cite{Ioka:2014kca, Ng:2014pca, Ibe:2014pja, DiFranzo:2015qea, Shoemaker:2015qul} predicted the effect of $\nu$SI off the C$\nu$B on the diffuse flux of high-energy astrophysical neutrinos, in the context of IceCube observations.  Reference\ \cite{Mohanty:2018cmq} studied the  case where the $\nu$SI spectral dip may explain the lack of IceCube events beyond a few PeV.  Ours is the first rigorous statistical analysis looking for $\nu$SI in the IceCube diffuse neutrino flux.  Below, we compare similarities and differences of our analysis to \Refe\ \cite{Ng:2014pca}, which is representative of the above works.

Reference\ \cite{Ng:2014pca} focused on a few benchmark choices of $M$ and $g$ and their compatibility with IceCube data.  Our best-fit values for the $\nu$SI parameters ($M = 14.26$~MeV and $g = 0.1$) are close to model B ($M = 10$~MeV and $g = 0.3$) of \Refe\ \cite{Ng:2014pca}.  Model B, like our best fit, was found to reproduce the IceCube event spectrum, in particular, the lack of observed between 300~TeV and 1~PeV.  

However, there are key differences between our work and \Refe\ \cite{Ng:2014pca}.  First, \Refe\ \cite{Ng:2014pca} assumed a hard spectral index of $\gamma = 2$, whereas we find $\gamma = 2.74$, because we start our fit at a lower deposited energy, \ie, 10~TeV vs.~100~TeV.  Second, we model the background of atmospheric neutrinos and muons in detail, including using an updated version of the IceCube self-veto.  Third, we account for the changes to the neutrino flux introduced during its propagation inside the Earth.  Fourth, we use both HESE showers and tracks, whereas \Refe\ \cite{Ng:2014pca} used only showers.  Fifth, we compute the event rate at IceCube differently for all flavors, accounting for their differences in deposited energy.  These improvements allow us to place robust limits on $\nu$SI, while remaining compatible with the predictions of \Refs\ \cite{Ioka:2014kca, Ng:2014pca, Ibe:2014pja, DiFranzo:2015qea, Shoemaker:2015qul}. 

In addition, \Refe\ \cite{Kelly:2018tyg} estimated the limits on $\nu$SI inferred from the recent observation of a burst of neutrinos possibly associated to the blazar TXS~0506+056\ \cite{IceCube:2018cha}; see \figu{limits}.  Unlike our limits, the limits from \Refe\ \cite{Kelly:2018tyg} were derived solely from the survival of the emitted neutrinos en route to Earth, not from looking for $\nu$SI-induced spectral features in the detected data.  

\subsection{Limitations and improvements}  

In our analysis, we fixed the unknown value of the neutrino mass to $m_\nu = 0.1$~eV and, with it, computed $E_{\rm res}$ for each tested value of $M$.  Using smaller values of $m_\nu$ would shift the limits in \figu{limits} to lower values of $M$.  A refined search for $\nu$SI could adopt the existing upper limits on the sum of the masses of all neutrino species\ \cite{Lattanzi:2017ubx} as informed priors on $m_\nu$, instead of fixing its value.

The best-fit values of $M$ and $g$ stem from attempting to fill the gap in events in the 6-year HESE sample.  However, the gap admits alternative explanations.  For instance, it could instead be due to the existence of multiple populations of sources, each producing neutrinos in different energy ranges\ \cite{Denton:2017jwk, Denton:2018tdj, Palladino:2017qda, Palladino:2018evm}.  Currently, a comparison of alternatives is of little interest, given the lack of statistically significant evidence for $\nu$SI. 

An interesting possibility that lies beyond the scope of this paper is to complement the use of HESE events with through-going muons, \ie, muon tracks born outside the detector that cross part of it\ \cite{Aartsen:2016xlq}.  Detected through-going muons number in the hundreds of thousands, though only a small fraction of them is publicly available.  On the one hand, including through-going muons would increase the statistics.  On the other hand, unlike HESE events, through-going muons have a high contamination of atmospheric neutrinos and muons, and a large uncertainty on the neutrino energy reconstruction, which might dilute any $\nu$SI-induced features in the neutrino energy spectrum.


\section{Summary and outlook}  
\label{section:conclusions}

For the first time, we have performed a rigorous statistical analysis in search for evidence of {\it secret neutrino interactions} ($\nu$SI), \ie, new neutrino-neutrino interactions that are not contained in the Standard Model, in the diffuse flux of TeV--PeV astrophysical neutrinos detected by the IceCube neutrino telescope.  

We modeled the propagation of high-energy neutrinos from their extragalactic sources to Earth, across distances of Mpc--Gpc, accounting for their undergoing $\nu$SI off the cosmic neutrino background along the way.   If $\nu$SI occur, they would leave a characteristic imprint: a relatively narrow deficit of events, or dip, in the spectrum of high-energy neutrinos.  

We looked for this dip in IceCube neutrino data.  We accounted in detail for the propagation of neutrinos inside Earth, which modifies the shape of the energy spectrum in a flavor- and direction-dependent manner, for their detection at IceCube, which dilutes the spectral features, and for the background of atmospheric neutrinos and muons, which muddles the astrophysical signal.

In 6 years of publicly available IceCube High Energy Starting Events (HESE)\ \cite{Kopper:2017zzm}, we found no statistically significant evidence of $\nu$SI.  Thus, we placed upper limits on the coupling strength of the new mediator through which $\nu$SI occurs in the mass range of 1--100~MeV.  Figure\ \ref{fig:limits} shows that our limits confirm existing limits derived using substantially different methods than ours.

Our limits are largely driven by the absence of HESE events detected between 300~TeV and 1~PeV in the 6-year sample.  In the future, if this gap in events remains in spite of growing statistics, it would build up the evidence for $\nu$SI.  However, preliminary 7.5-year HESE results\ \cite{Wandkowsky:2018} contain a few new events in this energy range; if confirmed, they would instead strengthen the limits on $\nu$SI.  In the future, using the proposed upgrade IceCube-Gen2\ \cite{Aartsen:2019swn}, with an event rate 5--7 times higher than IceCube, a significantly larger event sample would reinforce the outcome in either case.

\medskip


\section*{Acknowledgements}

We thank Carlos Arg\"uelles, John Beacom, Kevin Kelly, Jennifer Kile, Samuel McDermott, Kohta Murase, Kenny Ng, Sergio Palomares-Ruiz, and Austin Schneider for useful discussions.  This project was supported by the Villum Foundation (Project No.~13164), the Carlsberg Foundation (CF18-0183), the Knud H{\o}jgaard Foundation, and the Deutsche Forschungsgemeinschaft through Sonderforschungbereich SFB~1258 ``Neutrinos and Dark Matter in Astro- and Particle Physics'' (NDM).



%


\appendix


\section{High-energy neutrino detection}  
\label{appendix:detection}


\subsection{IceCube HESE events}
\label{section:contained}

In our analysis, we use IceCube High Energy Starting Events (HESE)\ \cite{Aartsen:2013jdh, Aartsen:2014gkd}, where the neutrino interaction occurs inside the detector and the outer layer of PMTs serves as a self-veto to reduce contamination from atmospheric neutrinos\ 
\cite{Schonert:2008is, Gaisser:2014bja, Arguelles:2018awr}.

In a neutrino-nucleon interaction (see Section\ \ref{section:prop_in_earth}), the final-state hadrons carry a fraction $y$, the inelasticity, of the parent neutrino energy, and the final-state lepton carries the remaining fraction $(1-y)$.  Final-state hadrons and charged leptons shower and radiate light.  At neutrino energies of TeV and up, the average value of the inelasticity is about 30\% \cite{Gandhi:1995tf, Gandhi:1998ri, Connolly:2011vc, CooperSarkar:2011pa}.  The higher the energy given to final-state charged particles, the higher the energy $E_{\rm dep}$ deposited in the detector.

HESE showers are fully contained in the detector, so $E_{\rm dep} = E_{\rm sh}$, where $E_{\rm sh}$ is the energy of the particles in the shower.  In CC interactions, the shower is due to the final-state hadrons and lepton, so $E_{\rm sh} = E_\nu$, while in NC interactions the shower is due only the final-state hadrons, so $E_{\rm sh} = y E_\nu$.  

HESE tracks start inside the detector, but are only partially contained by it.  Yet, the measured rate of energy loss of the muon as it propagates approximates the muon energy $E_\mu$.  Hence, $E_{\rm dep} = E_{\rm sh} + E_\mu \approx E_\nu$, where the shower is due to final-state hadrons only.


\subsection{Computing HESE neutrino rates}  

To compute the HESE event rate, we follow \Refe\ \cite{Palomares-Ruiz:2015mka}, which treats separately the contributions of different flavors, interaction channels, and decay channels of final-state particles, and accounts for the energy resolution of the detector.  Because the full procedure is elaborate, below we only sketch it and refer to \Refe\ \cite{Palomares-Ruiz:2015mka} for details.

We compute the energy distributions of HESE showers $dN^{{\rm sh}, f} / dE_{\rm dep}$ and tracks $dN^{{\rm tr}, f} / dE_{\rm dep}$ for
astrophysical neutrinos ($f = $~ast) and conventional atmospheric neutrinos ($f = $~atm), \ie, those coming from pion and kaon decays (see Appendix\ \ref{appendix:atm_bg}).  For each flux type $f$, the spectrum receives contributions from all flavors, \ie,
\begin{eqnarray}
 \label{equ:spectrum_total_sh}
 \frac{dN^{{\rm sh}, f}}{dE_{\rm dep}}
 &=&
 \sum_{\alpha=e,\mu,\tau} 
 \frac{dN_\alpha^{{\rm sh}, f}}{dE_{\rm dep}} \;,
 \\
 \label{equ:spectrum_total_tr}
 \frac{dN^{{\rm tr}, f}}{dE_{\rm dep}}
 &=&
 \sum_{\alpha=e,\mu,\tau} 
 \frac{dN_\alpha^{{\rm tr}, f}}{dE_{\rm dep}} \;.
\end{eqnarray}
The contribution of each flavor is:
\begin{eqnarray*}
 \frac{dN_\alpha^{{\rm sh}, f}}{dE_{\rm dep}}
 &=&
 \frac{dN_{\nu_\alpha}^{{\rm sh}, {\rm NC}, f}}{dE_{\rm dep}}
 +
 (1-\delta_{\alpha \mu})
 \frac{dN_{\nu_\alpha}^{{\rm sh}, {\rm CC}, f}}{dE_{\rm dep}}
 +
 \frac{dN_{\nu_\alpha}^{{\rm sh}, e, f}}{dE_{\rm dep}} 
 \\
 &&
 + \;
 (\nu_\alpha \to \bar{\nu}_\alpha)
 \;, \\
 \frac{dN_\alpha^{{\rm tr}, f}}{dE_{\rm dep}}
 &=&
 (1-\delta_{\alpha e})
 \frac{dN_{\nu_\alpha}^{{\rm tr}, {\rm CC}, f}}{dE_{\rm dep}}
 +
 \frac{dN_{\nu_\alpha}^{{\rm tr}, e, f}}{dE_{\rm dep}} 
 +
 (\nu_\alpha \to \bar{\nu}_\alpha)
 \;.
\end{eqnarray*}

The dominant detection channel is neutrino-nucleon interaction (see Section\ \ref{section:contained}), NC and CC.
We also include the contribution from neutrino-electron interaction ($e$).  It is sub-dominant except around $E_\nu \approx 6.3$~PeV, where the Glashow resonance $\bar{\nu}_e + e \to W$\ \cite{Glashow:1960zz} dominates.  The resonance does not affect our analysis significantly, since there are no events beyond 2~PeV in the 6-year HESE sample that we use (see Section \ref{section:testing_nusi}), but we account for it when computing the denominators of the probability distribution functions $\mathcal{P}_{i, \rm{ast}}$, $\mathcal{P}_{i, \rm{atm}}$, and $\mathcal{P}_{i, \mu}$ in Eqs.\ (\ref{equ:pdf_ast}) and (\ref{equ:pdf_conv}) in the main text.

As illustration, the shower spectrum due to NC interactions of $\nu_\alpha$ is
\begin{widetext}
 \begin{equation}
  \label{equ:spectrum_sh_nc}
  \frac{dN_{\nu_\alpha}^{{\rm sh}, {\rm NC}, f}(E_{\rm dep})}{dE_{\rm dep}}
  = 
  T N_{\rm A}
  \int_0^\infty
  dE_\nu
  \frac{d\phi_{\nu_\alpha}^f(E_\nu)}{dE_\nu}
  \int_0^1
  dy
  M_{\rm eff}(E_{\rm true}(E_\nu))
  R(E_{\rm true}(E_\nu), E_{\rm dep}, \sigma(E_{\rm true}(E_\nu))
  \frac{d\sigma_{\nu_\alpha}^{\rm NC}(E_\nu, y)}{dy} \;,
 \end{equation}
\end{widetext}
where $T = 6$~yr is the detector live time, $N_{\rm A} = 6.022 \times 10^{-23}$~g$^{-1}$ is Avogadro's number, $d\phi_{\nu_\alpha}^f / dE_\nu$ is the flux of $\nu_\alpha$ that reaches the detector after propagating inside Earth (for atmospheric neutrinos, it is the flux multiplied by the self-veto passing fraction; see Appendix\ \ref{appendix:atm_bg}), $M_{\rm eff}$ is the effective detector mass\ \cite{Palomares-Ruiz:2015mka}, and $\sigma_{\nu_\alpha}^{\rm NC}$ is the NC cross section.  The energy resolution function $R$ accounts for the mismatch between the measured deposited energy, $E_{\rm dep}$ and the true deposited energy, $E_{\rm true}$, which varies with $E_\nu$.  It is a Gaussian with a spread of $\sigma \approx 0.12 E_{\rm true}$\ \cite{Palomares-Ruiz:2015mka}.  The contribution of $\bar{\nu}_\alpha$ is the same as \equ{spectrum_sh_nc}, with $\nu_\alpha \to \bar{\nu}_\alpha$.  

For the contribution of CC interactions of $\nu_e$ and $\nu_\tau$ to the shower rate and of CC interactions of $\nu_\mu$ to the track rate, the expressions are  similar to \equ{spectrum_sh_nc}, with NC $\to$ CC.  For the contribution of CC interactions of $\nu_\tau$, the calculation is more complex.  Below a few PeV, the tauon created in the interaction decays inside the detector and the particular decay channel determines the deposited energy and whether the decay contributes to the shower rate or the track rate.  We compute the contribution of each tauon decay channel separately.  Similarly, we compute separately the contribution of each decay channel of the $W$ boson created in the Glashow resonance.  
In both cases, we follow \Refe\ \cite{Palomares-Ruiz:2015mka}.  
The relation between $E_{\rm true}$ and $E_\nu$ changes depending on the flavor, interaction channel, and decay channel\ \cite{Palomares-Ruiz:2015mka}.  

We compute the differential deep-inelastic-scattering cross sections on protons and neutrons\ \cite{Giunti:2007ry}, $d\sigma_{p, \nu_\alpha}^{\rm NC} / dy$, $d\sigma_{n, \nu_\alpha}^{\rm NC} / dy$, and their CC equivalents, using the recent CTEQ14 parton distribution functions\ \cite{Dulat:2015mca}, for $\nu_\alpha$ and $\bar{\nu}_\alpha$.  Because interactions occur in ice, we weight the cross sections by the mass number $A = 18$, atomic number $Z = 10$, and neutron number $N = 8$ of water, \ie,
\begin{equation*}
 \frac{d\sigma_{\nu_\alpha}^{\rm NC}}{dy}
 =
 \frac{1}{A}
 \left(
 Z
 \frac{d\sigma_{p, \nu_\alpha}^{\rm NC}}{dy}
 +
 N
 \frac{d\sigma_{n, \nu_\alpha}^{\rm NC}}{dy}
 \right) \;,
\end{equation*}
and similarly for CC interactions.  For neutrino-electron interactions, we compute the differential cross section following \Refs\ \cite{Mikaelian:1980vd, Gandhi:1995tf}.


\section{Atmospheric backgrounds}  
\label{appendix:atm_bg}

For atmospheric neutrinos, we use {\tt MCEq}\ \cite{Fedynitch:2015zma, MCEq} to compute the fluxes of neutrinos and anti-neutrinos of all flavors at the surface of the Earth, coming from pion and kaon decays, \ie, conventional atmospheric neutrinos.  We do not consider prompt atmospheric neutrinos, coming from charmed meson decays, since they remain unobserved\ \cite{Aartsen:2015knd}.  We adopt the default model choices used in IceCube analyses: for the parent cosmic-ray flux, we use the Hillas-Gaisser model (H3a)\ \cite{Gaisser:2011cc}; for the particle interaction model, we use SIBYLL2.3c\ \cite{Engel:2019dsg}; for the atmospheric density profile, we use the NRLMSISE-00 model at the South Pole\ \cite{Picone:2002}.  The resulting neutrino fluxes are different for each flavor of $\nu$ and $\bar{\nu}$, and vary with energy and arrival direction.  We average the fluxes between their summer-time and winter-time values.  

Second, we implement the HESE self-veto to reduce the contribution of atmospheric neutrinos to the event rate.  We use {\tt nuVeto}\ \cite{Arguelles:2018awr, nuVeto} to compute the energy-dependent fraction of atmospheric $\nu$ and $\bar{\nu}$ of each flavor that pass the veto, reach the fiducial volume of the detector, and ultimately contribute to the event rate in our analysis.  The veto is more efficient for $\nu_e$ than for $\nu_\mu$; there is no veto for $\nu_\tau$, whose flux is tiny.  Veto efficiency grows with energy: at 10~TeV, the $\nu_\mu$ flux is reduced by up to a factor 2 and at 100~TeV, by up to a factor of 100.

For atmospheric muons, there is an irreducible background that reaches the detector and leaves tracks.  In our analysis, we directly estimate the spectrum of tracks due to atmospheric muons, $dN^{{\rm tr}, \mu}/dE_{\rm dep}$, following the parametrization of \Refe\ \cite{Vincent:2016nut}, which was built using publicly available IceCube data\ \cite{Aartsen:2014gkd}.


\end{document}